\documentclass[aps,prd,twocolumn,superscriptaddress]{revtex4}
\usepackage{graphicx}
\usepackage{amsmath}
\usepackage{amssymb}
\usepackage{slashed}
\usepackage{latexsym}
\usepackage{epsfig}
\usepackage{amsbsy}
\usepackage{array}
\usepackage{amssymb}
\usepackage{setspace}
\usepackage{bm}
\usepackage{lipsum}
\usepackage{mathrsfs}
\usepackage{float}
\usepackage{color}
\usepackage[T1]{fontenc}
\usepackage{mathptmx}
\DeclareMathAlphabet{\mathcal}{OMS}{cmsy}{m}{n}
\DeclareSymbolFont{largesymbols}{OMX}{cmex}{m}{n}
\begin{document}

\title{On the stability of two-flavor and three-flavor quark matter in quark stars within the framework of NJL model}

\author{Qianyi Wang}
\email{wangqianyi07@hotmail.com}
\affiliation{Department of physics, Nanjing University, Nanjing 210093, China}
\author{Tong Zhao}
\email{zhao708@purdue.edu}
\affiliation{Department of physics, Nanjing University, Nanjing 210093, China}
\author{Hongshi Zong}
\email{zonghs@nju.edu.cn}
\affiliation{Department of physics, Nanjing University, Nanjing 210093, China}
\affiliation{Nanjing Proton Source Research and Design Center, Nanjing 210093, China}
\affiliation{Department of physics, Anhui Normal University, Wuhu, Anhui 241000, China}
\date{\today}

\begin{abstract}

Following our recently proposed self-consistent mean field approximation approach, we have done some researches on the chiral phase transition of strong interaction matter within the framework of Nambu-Jona-Lasinio (NJL) model. The chiral susceptibility and equation of state (EOS) are computed in this work for both two-flavor and three-flavor quark matter for contrast. The Pauli-Villars scheme, which can preserve gauge invariance, is used in this paper. Moreover, whether the three-flavor quark matter is more stable than the two-flavor quark matter or not in quark stars is discussed in this work. In our model, when the bag constant are the same, the two-flavor quark matter has a higher pressure than the three-flavor quark matter, which is different from what Witten proposed in his pioneering work.

\end{abstract}

\maketitle
\section{\label{sec:level1}Introduction}

It has long been believed that the quark stars or the materials in the core of neutron stars are made up of strange quark matter, with comparable amounts of u, d and s quarks. Some pioneering discussions can be found in Refs. \cite{ivanenko1965hypothesis,ivanenko1969remarks,itoh1970hydrostatic,bodmer1971collapsed,chin1979possible,witten1984cosmic,farhi1984strange}. With the lack of a first-principle understanding of the strong interaction dynamics, these discussions are based on the MIT bag model. For example, as Witten argues in his work \cite{witten}, when exerting the same pressure, the Fermi momentum of the three-flavor quark matter is lower than the two-flavor quark matter. In 1999, studies on the quark matter in compact stars based on NJL-type models are performed \cite{schertler1999neutron,buballa1999strange}. Again,  following the Witten's conjecture, only three-flavor NJL model is considered in these works. However, in a recent study, the authors find that two-flavor quark matter generally has lower energy per baryon than normal nuclei and strange quark matter according to their model \cite{holdom2018quark}. This result suggests that the stability of quark matter is model dependent, and the interaction plays an important role.

On the other hand, as is pointed out in the Refs. \cite{Wang2019New,zhao2019does}, there exists a contradiction between the picture of hadron degree of freedom and the picture of quark degree of freedom. From the picture of quark degree of freedom, the system will undergo the well-known chiral phase transition along with the continuous increase of quark chemical potential to a critical value $\mu_c$. In NJL-type models, the value of $\mu_c$ is 330-380 MeV \cite{Cui2018Wigner,Masayuki1990Chiral,Buballa2004NJL}. However, from the picture of hadron degree of freedom, the quark matter begin to form when the density is larger than $4n_0$ according to percolation theory \cite{CELIK1980128}. $n_0$ represents the nuclear saturation density. The quark chemical potential corresponding to 4$n_0$ is about 580 MeV by the method of RMF with model parameters NL3$\omega\rho$ from the hadron degrees of freedom. To solve this contradiction, a new self-consistent method of mean field approximation is developed in Refs. \cite{Wang2019New,zhao2019does,wang2019non,yang2019qcd,li2020strange} to calculate the phase diagram of quark matter and the mass-radii relations of quark stars. In that model, the critical chemical potential $\mu_c$ of the chiral phase transition can be greatly increased by changing the parameter $\alpha$ which reflects the weight of different interaction channels in the model. However, only two-flavor NJL model is discussed and the three momentum cutoff regularization is adopted in their model for simplicity. That means the model is valid only if one assumes that the cutoff involved is much larger than all relevant momenta. So, the quark chemical potential cannot be greater than this cutoff, which will place a restriction on the maximal quark star mass based on the NJL model. 

As an expansion of our previous work \cite{zhao2019does}, both two-flavor and three-flavor quark matter are explored in this work based on the NJL model. While the three momentum cutoff scheme is widely used in the NJL-type models, we use the Pauli-Villars scheme that can preserve gauge invariance in this paper. Especially, whether the three-flavor quark matter is more stable than the two-flavor quark matter in quark stars is discussed. 

This paper is organized as follows: In Sec. \ref{sec:model}, we briefly introduce our self-consistent mean field theory of the NJL model, and work out the chiral susceptibility to specify the chiral phase transition point, then obtain the phase diagram for both two-flavor and three-flavor NJL models.  In Sec. \ref{sec:EOS}, the equations of state of two-flavor and three-flavor NJL models are derived to discuss the stability of quark matter. The mass-radii relations of quark stars are also computed in this section to show the stiffness of the equation of state. Sec. \ref{sec:discussion} is the summary and discussion of our work.

\section{\label{sec:model}NJL MODEL AND PHASE DIAGRAM}
NJL model is firstly developed by Nambu and Jona-Lasinio in 1961 to model spontaneous breaking of chiral symmetry in the vacuum in analogy with superconductivity \cite{nambu1995dynamical,nambu1961dynamical}. Then, in 1974, Eguchi and Sugawara introduced a two-flavor quark NJL model with up and down quarks \cite{eguchi1974extended}. In 1976, Kikawa generalized this model to three-flavor case \cite{kikkawa1976quantum}. The NJL model cannot be solved strictly, so the mean field approximation is widely used in the studies of NJL model. However, in the original mean field approximation, the Fierz transformation is not performed. To get a self-consistent result in the sense of mean field approximation, the contribution from the Fierz transformed Lagrangian must be taken into account \cite{kunihiro1984self,Hatsuda1985}. After that, the authors of Ref. \cite{RevModPhys.64.649} combined both the original Lagrangian ($\mathcal{L}$) and the Fierz identity of it ($\mathcal{L}_F$) to construct an equivalent Lagrangian as described below:
\begin{equation}
\mathcal{L}_{\mathrm{R}}=(1-\alpha)\mathcal{L}+\alpha\mathcal{L}_{F},
\end{equation}
where $\alpha=0.5$. In the effective Lagrangian the leading order term $O(N_c^0)$ of the large $N_c$ expansion and the next leading order term $O(\frac{1}{N_c})$ (introduced by Fierz transformation) are mixed to guarantee that the calculation of NJL model is theoretically self-consistent at the level of the mean field approximation \cite{RevModPhys.64.649}. Nonetheless, as shown in Refs. \cite{zhao2019does,Wang2019New}, there is no physical requirement that the value of $\alpha$ must be 0.5. So, in principle the value of $\alpha$ should be determined by experiments. Hence one can see that our new self-consistent mean field approximation differs from the previous mean field approximation only in the value of $\alpha$ \cite{zhao2019does,Wang2019New}. It should be noted that the Fierz transformation is a mathematically equivalent transformation and $\alpha$ can be any complex number in principle. However, there is a basic requirement for physical Lagrangian, that is, each term of Lagrangian must be Hermitian. So, we only treat $\alpha$ as a real number in this paper. In this manuscript, we first study the relationship of $\alpha$ and the stiffness of EOS. Then we will constrain $\alpha$ by astronomical observations of neutron stars. 

What we need to point out here is that the strange quark matter hypothesis can be traced back to the mid-eighties of the last century. Various consequences of the hypothesis for nucleon, particle and astrophysics have been explored ever since, but the hypothesis has neither been proven nor disproven. One of the main motivations of this paper is to discuss whether the three-flavor quark matter is more stable than the two-flavor quark matter in nature. The results of two-flavor model and three-flavor model are shown in Part A and Part B of this section respectively.
\subsection{\label{subsec:two flavor}TWO-FLAVOR NJL MODEL}
The standard two-flavor NJL model Lagrangian with interaction terms in scalar and pseudoscalar channel is given by
\begin{equation}
\mathcal{L}=\overline{\psi} \left(i\slashed{\partial}-m\right) \psi+G\left[(\overline{\psi} \psi)^{2}+(\overline{\psi} i \gamma^{5}\psi)^{2}\right],
\end{equation}
where $m$ is current quark mass and $G$ denotes the coupling constant of four-fermion interaction. The Fierz identity of the Lagrangian is
\begin{equation}
\begin{aligned}
\mathcal{L}_{F}=&\overline{\psi}(i\slashed{\partial}-m)\psi\\+&\frac{G}{8 N_c}\big[ 2(\overline{\psi} \psi)^{2}+2(\overline{\psi} i \gamma^{5} \tau \psi)^{2}-2(\overline{\psi} \tau \psi)^{2}\\-&2(\overline{\psi} i \gamma^{5} \psi)^{2} -4(\overline{\psi} \gamma^{\mu} \psi)^{2}-4(\overline{\psi} i \gamma^{\mu} \gamma^{5} \psi)^{2}\\+&(\overline{\psi} \sigma^{\mu \nu} \psi)^{2}-(\overline{\psi} \sigma^{\mu \nu} \tau \psi)^{2}\big],
\end{aligned}
\end{equation}
where $N_c$ is the number of color. It is interesting to compare Eq. (2) and Eq. (3). As is shown in Eq. (2), the original Lagrangian contains only the interactions of scalar and pseudoscalar channels. With the help of the Fierz transformation, all the hidden channels of original Lagrangian are released. In Eq. (3), the Fierz transformed Lagrangian contains not only the scalar and pseudoscalar interactions, but also vector and axialvector interaction channels. As is shown in Ref. \cite{zhao2019does}, the introduction of vector channel is very important when dealing with finite chemical potential and the axialvector channel plays an important role in the treatment of chiral chemical potential according to Ref. \cite{yang2019qcd}.

\begin{figure}
	\includegraphics[scale=0.96]{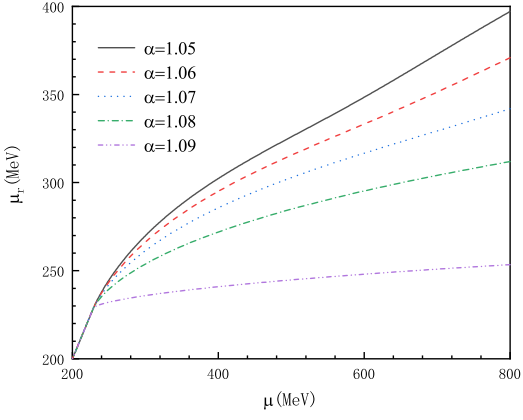}
	\caption{Effective chemical potential as functions of chemical potential of u quark with different $\alpha$ in the vicinity of $\frac{12}{11}$.}
	\label{fig:potential}
\end{figure}

\begin{figure}
	\includegraphics[scale=0.96]{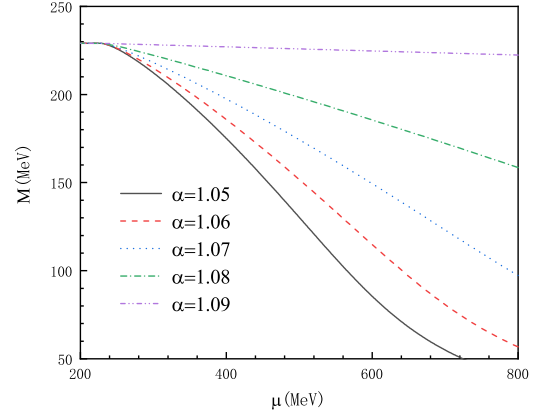}
	\caption{Effective mass as functions of chemical potential of u quark with different $\alpha$ in the vicinity of $\frac{12}{11}$.}
	\label{fig:mass}
\end{figure}

\begin{figure}
	\includegraphics[scale=0.96]{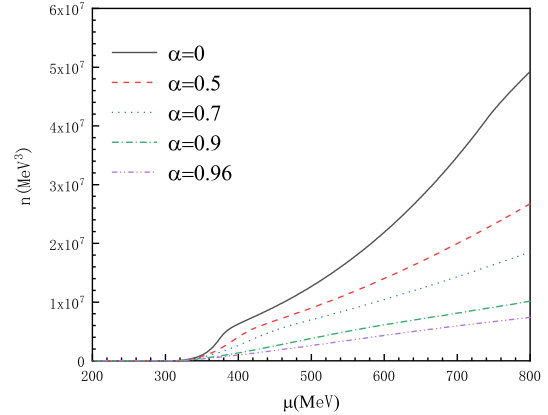}
	\caption{The quark number density as functions of chemical potential at zero temperature (two flavor).}
	\label{fig:n_u_10_2fla}
\end{figure}

\begin{figure}
	\includegraphics[scale=0.96]{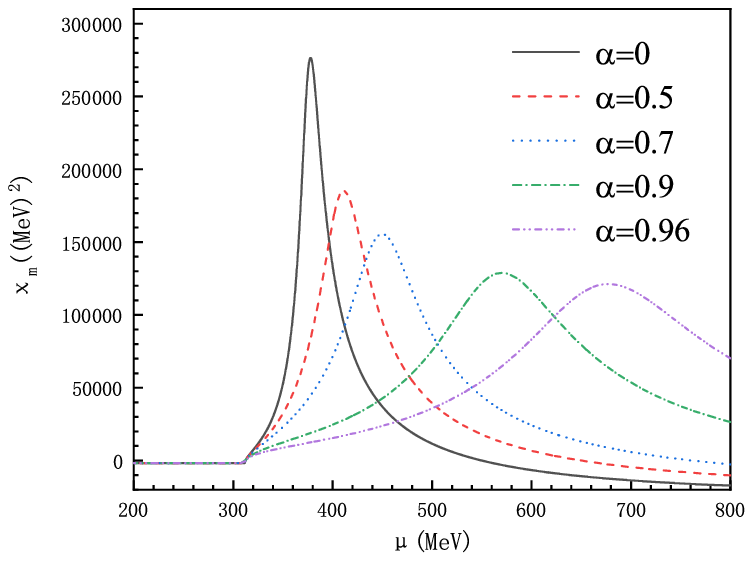}
	\caption{The chiral susceptibility as functions of chemical potential at zero temperature (two flavor).}
	\label{fig:x_u_10_2fla}
\end{figure}

\begin{figure}
	\includegraphics[scale=0.96]{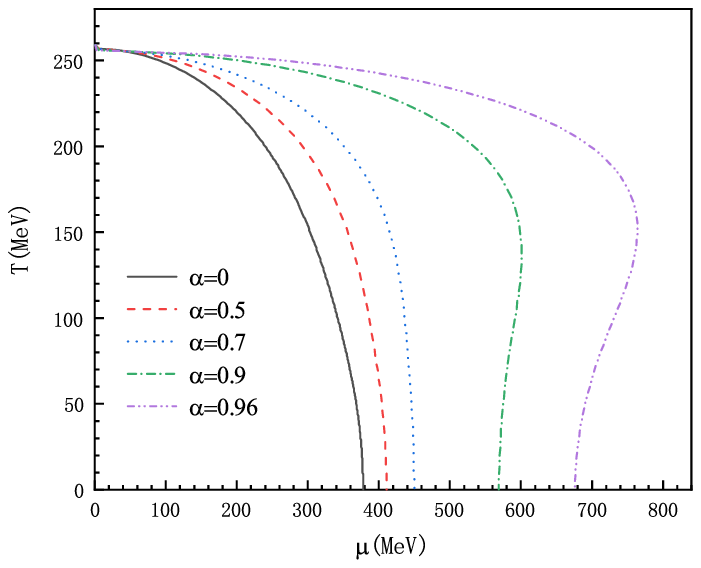}
	\caption{The phase diagram of strongly interacting matter with different parameter $\alpha$ (two flavor).}
	\label{fig:phase_diagram_10_2fla}
\end{figure}

Now we construct the most general equivalent Lagrangian and use the mean field approximation:
\begin{equation}
\begin{aligned}
	\mathcal{L}_{eff}=&(1-\alpha)\langle\mathcal{L}\rangle_{MFA}+\alpha \langle\mathcal{L}_{\mathrm{F}}  \rangle_{MFA}\\
	=&\overline{\psi} (i\slashed{\partial}-m)\psi+2(1-\alpha)G(\langle\overline{\psi} \psi\rangle\overline{\psi} \psi)\\
	&+\frac{\alpha G}{2N_c}\overline{\psi}(\langle\overline{\psi} \psi\rangle -2\langle\psi^+\psi\rangle\gamma^0)\psi+constant,\\
\end{aligned}
\end{equation}
where $\langle\quad\rangle_{MFA}$ represents applying mean field approximation and $\langle\quad\rangle$ means getting the expectation value of vacuum.
Here, we would like to point out that the combination of the original Lagrangian and the Fierz transformed Lagrangian indicates not only the contribution of the leading order term of the large $N_c$ expansion but also the next leading term is considered in our model. To simplify the expression of gap equation and keep it same as the usual one widely used, we introduce a new coupling constant $g = \left( 1-\frac{11}{12}\alpha \right)G$. Then the gap equation and expression of effective chemical potential can be expressed as:
\begin{equation}
\begin{aligned}
\mathrm{M} = &\mathrm{m}-2(1-\frac{11}{12}\alpha)G\sum_{f=u,d}\langle\overline{\psi_f} \psi_f\rangle\\
 = &\mathrm{m}-2g\sum_{f=u,d}\langle\overline{\psi_f} \psi_f\rangle,
\end{aligned}
\end{equation}
\begin{equation}
\begin{aligned}
\mu' = &\mu-\frac{\alpha}{3}G\sum_{f=u,d}\langle\psi^{+}_f \psi_f\rangle\\
=&\mu-\frac{\alpha}{3}\frac{12g}{12-11\alpha}\sum_{f=u,d}\langle\psi^{+}_f \psi_f\rangle,
\end{aligned}
\end{equation}
where $f = u, d$, and $m$ is the current quark mass. The value of $g$ can be derived by fitting the data of mass and decay constant of pion and the result is shown in Tab. \ref{table:two flavor parameter}. Although the Lagrangian in our model is equivalent to the standard one no matter how much the value of $\alpha$ is, after the mean field approximation, the vector-isoscalar interaction channel is introduced by $\alpha$, so the stiffness of the equation of state will increase with the growth of $\alpha$. However, the value of $\alpha$ cannot be unboundedly increased otherwise the sound speed of the equation of state will exceed the speed of light. Let us now discuss the possible range of value for $\alpha$: from the right hand of Eq. (6), it seems that when $\alpha=\frac{12}{11}$, the denominator of Eq. (6) goes to zero, but in fact the numerator of Eq. (6) also has the same factor that tends to zero at the same time. So the effective chemical potential is not destined to be infinity when $\alpha$ equals $\frac{12}{11}$. Our numerical results show that the effective chemical potential and effective quark mass will approach a certain value when $\alpha=\frac{12}{11}$. The behaviors of effective chemical potential and effective quark mass are shown in Fig. \ref{fig:potential} and Fig. \ref{fig:mass}. When $\alpha$ is greater than $\frac{12}{11}$, our algorithm fails to give a reliable solution of the gap equation. It needs to be clearly pointed out that the failure to find a convergent iterative solution for an integral equation does not mean the equation is really unsolvable. Therefore, in this paper, we use $\alpha=\frac{12}{11}$ as the upper limit of $\alpha$. Also, we want to emphasize that different from other models where extra interaction channels are added to the original NJL Lagrangian artificially, we construct an equivalent Lagrangian by Fierz transformation to reflect all the hidden possible interaction channels and protect the original NJL model from being damaged by factitiously introduced interaction terms \cite{yang2019qcd}.
It is well known that the NJL model is not renormalizable and the three-momentum cutoff is used as the regularization scheme in most studies. This regularization scheme will place a restriction on the quark chemical potential and thus place a restriction on the maximal quark star mass based on the NJL model \cite{zhao2019does}. So, the Pauli-Villars regularization is adopted in this paper. Pauli-Villars regularization can preserve gauge invariance and also leads to a higher maximal quark star mass. In the framework of Pauli-Villars regularization, the quark condensate of flavor $f$ is \cite{Meissner1990Regularization}:
\begin{equation}
\begin{aligned}
&\langle\overline{\psi}_f \psi_f\rangle=-4M_fN_c \int \frac{d^{4} \xi}{(2 \pi)^{4}} \frac{1}{\xi^{2}-M_f^{2}} \\ & \rightarrow-4 M_f N_{c}N_f \int \frac{d^{4} \xi}{(2 \pi)^{4}}\left(\frac{1}{\xi^{2}-M_f^{2}}-\frac{a_{1}}{\xi^{2}-\Lambda_{1}^{2}}-\frac{a_{2}}{\xi^{2}-\Lambda_{2}^{2}}\right),
\end{aligned}
\end{equation}
where $f=(u,d), \xi=(\xi_0,\vec{\xi})=(p_0+\mu',\vec{p})$ ($\mu'$ is effective chemical potential), $\Lambda_{1}$ and $\Lambda_{2}$ are cutoff parameters. $N_c$ represents the number of color and $M_f$ is the effective quark mass of flavor $f$ in the gap equation. By selecting appropriate values of $a_1$ and $ a_2$ we can make quark condensate finite. The values are listed as
\begin{equation}
\left\{\begin{array}{l}{a_{1}=\frac{M^{2}-\Lambda_{2}^{2}}{\Lambda_{1}^{2}-\Lambda_{2}^{2}}}, \\ {a_{2}=\frac{\Lambda_{1}^{2}-M^{2}}{\Lambda_{1}^{2}-\Lambda_{2}^{2}}}.\end{array}\right.
\end{equation}
For simplicity, we can set the cutoff scales $\Lambda_{1}=\Lambda_{2}=\Lambda$ after the subtraction. So only one cutoff parameter is left. In the finite temperature case, the final expression becomes:
\begin{equation}
\begin{aligned}
\langle\overline{\psi}_f \psi_f \rangle
=&-2MN_c \int \frac{d^{3} p}{(2 \pi)^{3}}\left[\frac{1}{E_{p}}(1-f-\overline{f})-\frac{3 E_{\Lambda}^{2}-E_{p}^{2}}{2 E_{\Lambda}^{3}}\right]\\
&+M N_c\int  \frac{d^{3} p}{(2 \pi)^{3}} \frac{E_{p}^{2}  -3 E_{\Lambda}^{2} }{E_{\Lambda}^{3} }\left(f_{\Lambda}+\overline{f_{\Lambda}}\right)\\
&+M N_c \int  \frac{d^{3} p}{(2 \pi)^{3}}\frac{E_{\Lambda} E_{p}^{2}-E_{\Lambda}^{3}}{E_{\Lambda}^{3} T}\left(f_{\Lambda}+\overline{f_{\Lambda}}\right)\\
&-MN_c \int \frac{d^{3} p}{(2 \pi)^{3}} \frac{\Lambda^{2}-M^{2}}{E_{\Lambda}^{2} T}(f_{\Lambda}^{2}+\overline{f_{\Lambda}}^{2}).
\end{aligned}
\end{equation}
Based on the same method, the number density can be obtained:
\begin{equation}
	 \langle\psi^{+}_f \psi_f \rangle =2 N_c \int \frac{d^{3} p}{(2 \pi)^{3}}\left[(f-\overline{f})-\left(f_{\Lambda}-\overline{f_{\Lambda}}\right)\right],
\end{equation}
where
\begin{equation}
f_{\Lambda}(p, \mu')=\frac{1}{1+e^{\frac{E_{\Lambda}-\mu'}{T}}},
\end{equation}
\begin{equation}
\overline{f_{\Lambda}}(p, \mu')=\frac{1}{1+e^{\frac{E_{\Lambda}+\mu'}{T}}},
\end{equation}
\begin{equation}
f(p, \mu')=\frac{1}{1+e^{\frac{E-\mu'}{T}}},
\end{equation}
\begin{equation}
\overline{f}(p, \mu')=\frac{1}{1+e^{\frac{E+\mu'}{T}}}.
\end{equation}
In these equations,  $E=\sqrt{\vec{p}^{2}+M^{2}}$ and $E_{\Lambda}=\sqrt{\vec{p}^{2}+\Lambda^{2}}$.
We first study the chiral phase transition under different values of $\alpha$ at zero temperature. The parameters we choose is listed in Table \ref{table:two flavor parameter}.
To determine the order of phase transition and the location of it, the chiral
susceptibility is also computed \cite{Cui2015Progress}.
\begin{equation}
\chi_{m}=\frac{\partial}{\partial m}\langle\overline{\psi} \psi\rangle.
\end{equation}
\begin{table}
	\caption{Parameters in Pauli-Villars regularization (two flavor). $m_\pi=135~\mathrm{MeV}$ and $f_\pi=94~\mathrm{MeV}$ represent the mass and the decay constant of pion respectively and the following parameters are derived by fitting them \cite{Inagaki2015Regularization}.}
	\label{table:two flavor parameter}
	\begin{ruledtabular}
		\begin{tabular}{ccccc}
			\textrm{$m_u$ (MeV)}&
			\textrm{$\Lambda$ (MeV)}&
			\textrm{$g ({\rm MeV}^{-2})$}\\
			\colrule
			10.0&778&$9.64\times10^{-6}$
		\end{tabular}
	\end{ruledtabular}
\end{table}
We assume the chemical potentials and current quark masses of u and d quarks are equal to each other, so the quark number densities and chiral susceptibilities of u and d quarks are also the same. The quark number densities and chiral susceptibilities with different $\alpha$ are shown respectively in Fig. \ref{fig:n_u_10_2fla} and Fig. \ref{fig:x_u_10_2fla}.
It can be seen from both Fig. \ref{fig:n_u_10_2fla} and Fig. \ref{fig:x_u_10_2fla}, the quark number densities and chiral susceptibilities are non-trivial when the quark chemical potential is greater than around 308 MeV. This is a model-independent result that quarks can be excited from the vacuum only if the chemical potential is comparable with the constituent quark mass according to Ref. \cite{halasz1998phase}. From Fig. \ref{fig:x_u_10_2fla} we can see that the chiral phase transition is a crossover under the Pauli-Villars regularization scheme no matter how large the $\alpha$ is, which is different from the three-cutoff situation \cite{Wang2019New}. That means there is no critical end point (CEP) according to our model. Although the CEP exists in most NJL-type models with a three-momentum cutoff regularization scheme \cite{xu2018qcd,ferreira2018presence}. Ref. \cite{cui2017proper} shows that the CEP can disappear if another regularization scheme is implemented. Our result is in accordance with Refs. \cite{cui2017proper,Inagaki2015Regularization}. Besides, the phase transition point can be remarkably postponed by increasing the parameter $\alpha$ in our model. The phase transition is the result of the competition between different interaction channels, and the parameter $\alpha$ in our model indicates the weights of different interaction channels. So, if we change the value of $\alpha$, the phase transition point will move accordingly. Here we would like to point out that although from the Lagrangian it seems that only the vector-isoscalar channel plays an important role, our model are different from those models where a vector-isoscalar term is added factitiously, because in physics the original NJL Lagrangian is damaged if there are other terms added by hand. The phase diagram is Fig. \ref{fig:phase_diagram_10_2fla}. It is noteworthy that there exist positive slope ($\frac{dT_c}{d\mu_c}>0$) in our phase diagram. According to Ref. \cite{PhysRevD.97.034022}, the positive slope in phase diagram coincides with negative entropy when calculated in quark-meson model with functional Renormalization Group (FRG) approach used. However, we find no negative entropy in our phase diagrams which are computed within the framework of NJL model and mean filed approximation.
\subsection{\label{subsec:three flavor}THREE-FLAVOR NJL MODEL}
The Lagrangian of three-flavor NJL model is
\begin{equation}
	\begin{aligned}
	\mathcal{L}_{\mathrm{NJL}} &=\overline{\psi}\left(i \slashed{\partial}-m\right) \psi\\&+G \sum_{a=0}^{8}\left[\left(\overline{\psi} \lambda^{a} \psi\right)^{2}+\left(\overline{\psi} i \gamma^{5} \lambda^{a} \psi\right)^{2}\right] \\ &-K\left[\operatorname{det} \overline{\psi}\left(1+\gamma^{5}\right) \psi+\operatorname{det} \overline{\psi}\left(1-\gamma^{5}\right) \psi\right].
	 \end{aligned}
\end{equation}
Where $m$ is the current quark mass and $G$ and $K$ denote the coupling constant. $\lambda^a$ is SU(3) generator in flavor space and $\lambda^{0} \equiv \sqrt{\frac{2}{3}} \mathbf{1}$. The determinant is also in flavor space.
The Fierz identity of it is :
\begin{equation}
\begin{aligned}
	\mathcal{L}_{F} &=\psi\bar(i\slashed{\partial}-m)\psi \\
	&-G\frac{1}{2}\sum_{i=0}^{8}\left[\left(\overline{\psi} \gamma^{\mu} \lambda^{C}_i \psi\right)^{2}-\left(\overline{\psi} \gamma^{\mu} \gamma^{5} \lambda^{C}_i \psi\right)^{2}\right] \\
	&-K\left[\operatorname{det} \overline{\psi}\left(1+\gamma^{5}\right) \psi+\operatorname{det} \overline{\psi}\left(1-\gamma^{5}\right) \psi\right].
\end{aligned}
\end{equation}
$\lambda^C$ is the SU(3) generator in color space and $\lambda_{0}^C \equiv \sqrt{\frac{2}{3}} \mathbf{1}$. We just take color singlet into consideration to simply the computation , so in transformed four-fermion interaction, only the term with $\lambda_0^C$ will be reserved. Apply new self-consistent mean field approximation to three-flavor NJL model, we can get the final Lagrangian:
\begin{equation}
\begin{aligned}
	\mathcal{L}_{\rm R}=&\overline{\psi}(i\slashed{\partial}-m_0)\psi \\
	+&(1-\alpha) G \sum_{a=0}^{8}\left[\left(\overline{\psi} \lambda^{a} \psi\right)^{2}+\left(\overline{\psi} i \gamma^{5} \lambda^{a} \psi\right)^{2}\right] \\
	-&\alpha G\frac{1}{2}\sum_{i=0}^8\left[\left(\overline{\psi} \gamma^{\mu} \lambda^{C}_i \psi\right)^{2}-\left(\overline{\psi} \gamma^{\mu} \gamma^{5} \lambda^{C}_i \psi\right)^{2}\right] \\
	-&K\left[\operatorname{det} \overline{\psi}\left(1+\gamma^{5}\right) \psi+\operatorname{det} \overline{\psi}\left(1-\gamma^{5}\right) \psi\right].
\end{aligned}
\end{equation}
It is noteworthy that we only consider part of the Fierz identity of three-flavor model. The contribution of Fierz identity of six-fermion interaction term to gap equation is the modification of coupling constant of $K$ and it has nothing to do with effective chemical potential. Because $K$ needs to be determined by parameter fitting under vanishing chemical potential, the modification has no effect on the value of $K$ when we do computation. So, we don't need to consider this effect of it and just apply Fierz transformation to four-fermion interaction term only. More details can be found in Ref. \cite{li2020strange}. Apply mean field approximation to the Lagrangian and the only non-vanishing terms will be $\langle\overline{\psi} \psi\rangle $and $\langle\overline{\psi} \gamma^0 \psi\rangle$. The gap equations are:
\begin{equation}
\begin{cases}
M_{u} &= m_{u}-4 g\sigma_u+2 K\sigma_d\sigma_s,\\
M_{d} &= m_{d}-4 g\sigma_d+2 K\sigma_u\sigma_s,\\
M_{s} &= m_{s}-4 g\sigma_s+2 K\sigma_u\sigma_d,
\end{cases}
\end{equation}
where $M_f (f=u,d,s)$ are constituent quark mass, $g=(1-\alpha)G, \sigma_u=\langle\overline{\psi_u}\psi_u\rangle,\sigma_d=\langle\overline{\psi_d}\psi_d\rangle,\sigma_s=\langle\overline{\psi_s}\psi_s\rangle$. The expressions of effective chemical potential are
\begin{equation}
\mu'_{f}=\mu_{f}-\frac{2}{3} \frac{g\alpha}{1-\alpha} \sum_{f'=u,d,s}\langle\psi_{f'}^{+} \psi_{f'}\rangle.
\end{equation}
\begin{figure}
	\includegraphics[scale=0.96]{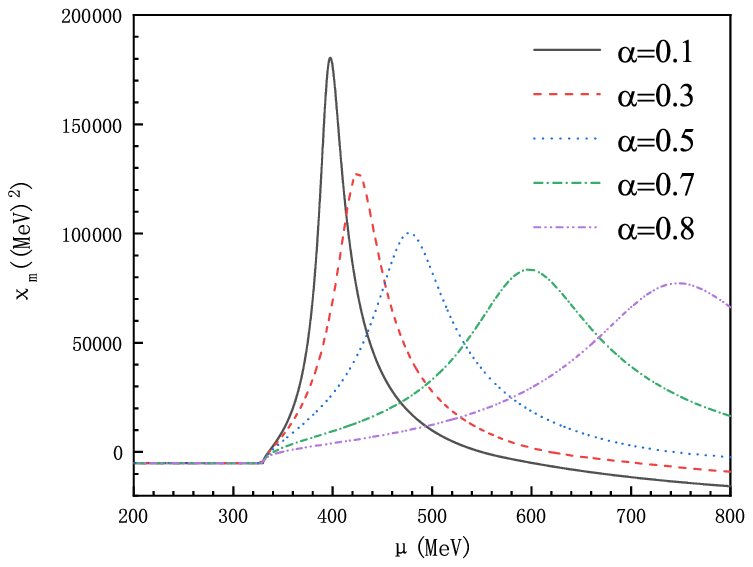}
	\caption{The chiral susceptibility as functions of chemical potential at zero temperature (three flavor).}
	\label{fig:x_u_3fla}
\end{figure}
\begin{figure}
	\includegraphics[scale=0.96]{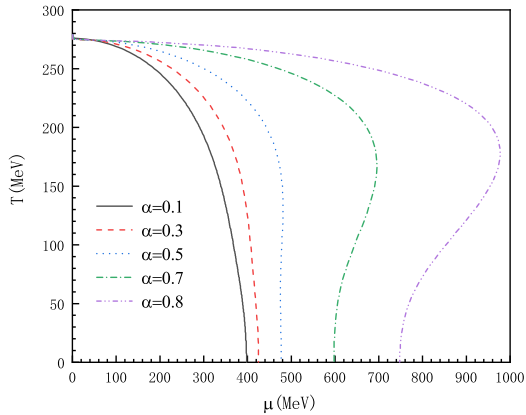}
	\caption{The phase diagram of strongly interacting matter with different parameter $\alpha$ (three flavor).}
	\label{fig:phase_diagram_3fla}
\end{figure}
The parameters are listed in Table \ref{table:three flavor parameter}. Just like what we do for two-flavor condition, we assume that the chemical potentials of u,d and s quark are equal to each other. Then we could derive the susceptibility at different chemical potentials of u quark, see Fig. \ref{fig:x_u_3fla}. The phase transition keeps crossover as two-flavor condition.
\begin{table}
	\caption{Parameters in Pauli-Villars regularization (three flavor) \cite{Kohyama2016Parameter}. $m_\pi=138~\mathrm{MeV}$ and $f_\pi=92~\mathrm{MeV}$ denote the mass and decay constant of pion respectively. $m_K=495~\mathrm{MeV}$ represents the mass of K and $m_{\eta'}=958~ \mathrm{MeV}$ is the mass of $\eta'$. The following parameters are derived by fitting them.}
	\label{table:three flavor parameter}
	\begin{ruledtabular}
		\begin{tabular}{ccccc}
			\textrm{$m_u$ (MeV)}&
			\textrm{$m_s$ (MeV)}&
			\textrm{$\Lambda$ (MeV)}&
			\textrm{$g\Lambda^2$}&
			\textrm{$K\Lambda^5$}\\
			\colrule
			11.8&327.8&743.3&5.885&175.5
		\end{tabular}
	\end{ruledtabular}
\end{table}
Carry on the steps we have done in two-flavor condition and connect the $\mu_c$ at different temperature, we could draw the phase diagram of quark matter (see Fig. \ref{fig:phase_diagram_3fla}).

\section{\label{sec:EOS}THE EQUATION OF STATE AND MASS-RADII RELATIONS}
To calculate the EOS for quark star model, we take chemical equilibrium into account. The chemical equilibrium and electric neutrality for two-flavor quark matter are
\begin{equation}
\left
\{\begin{array}
{l}{\mu_{d}=\mu_{u}+\mu_{e}}, \\
 {\frac{2}{3}n_u=\frac{1}{3}n_d+n_e}.
\end{array}
\right.
\end{equation}
The chemical equilibrium and electric neutrality for three-flavor quark matter are 
\begin{equation}
\left
\{\begin{array}
{l}{\mu_{d}=\mu_{u}+\mu_{e}}, \\
 {\mu_{d}=\mu_{s}},\\
 {\frac{2}{3} n_{u}=\frac{1}{3} n_{d}+\frac{1}{3} n_{s}+n_{e}}.
\end{array}
\right.
\end{equation}
From the expression of quark number densities, the pressure and energy densities can be obtained by \cite{zong2008calculation,zong2008model,yan2012connecting,benvenuto1995strange}
\begin{equation}
P=P(\mu=0, M)+\int_0^{\mu_u}\rho(\mu')d{\mu'},
\end{equation}
\begin{equation}
\epsilon=-P+\sum_i\mu_i\rho_i,
\end{equation}
where $P\left(\mu=0, M\right)$ denotes the vacuum pressure, which is a density-independent quantity and $M$ is a solution of the quark gap equation. It is usually associated with the bag constant $P\left(\mu=0, M\right)=-B$. Although the bag constant can be calculated from the NJL model, the results depend on the model and parameters, and there is no reliable way from the first principle of QCD to calculate the value of the vacuum pressure. Detailed discussion on the bag constant can also be found in our previous work Ref. \cite{zhao2019does}. The bag constant has an empirical domain which ranges from $(100 {\rm MeV})^4$ to $(200 {\rm MeV})^4$ \cite{Gao1992,lu1998medium}. So, it is considered as a free parameter in this domain in this paper. 

\begin{figure}
	\includegraphics[scale=0.96]{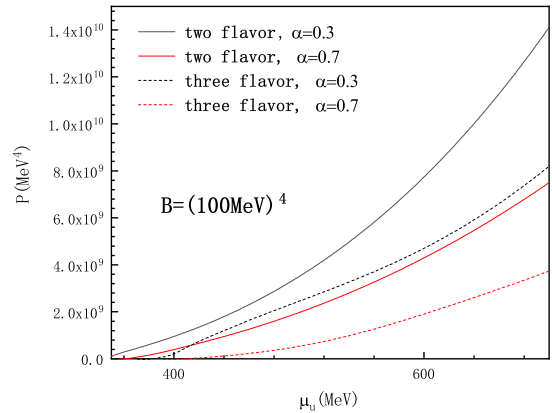}
	\caption{The pressure as functions of chemical potential for two-flavor and three-flavor conditions at different value of $\alpha$.}
	\label{fig:compare}
\end{figure}

\begin{figure}
	\includegraphics[scale=0.96]{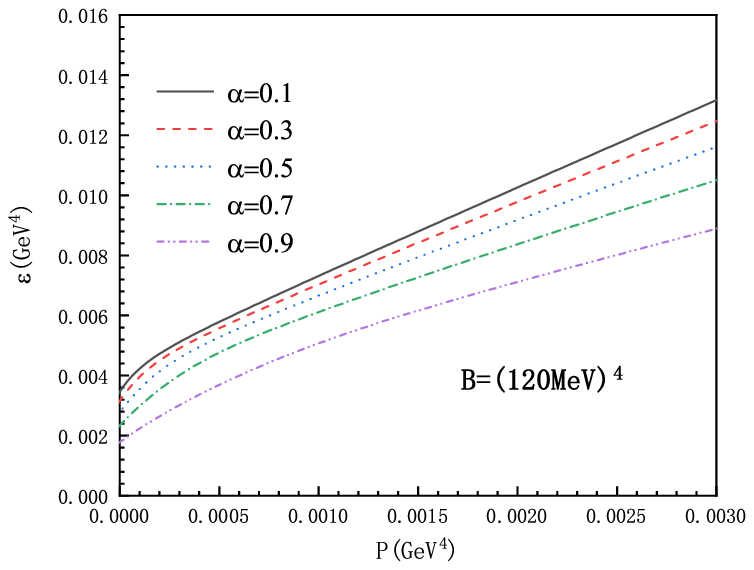}
	\caption{Several EOSs with $\alpha$ = 0.1, 0.3, 0.5, 0.7, 0.8 (two flavor). With a fixed negative pressure of vacuum: $B$= (100 ${\rm MeV})^4$, the stiffness of EOS increases along with $\alpha$.}
	\label{fig:EOS_2fla}
\end{figure}

\begin{figure}
	\includegraphics[scale=0.96]{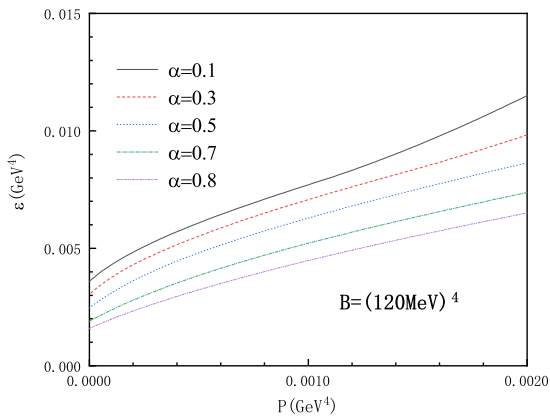}
	\caption{Several EOSs with $\alpha$ = 0.1, 0.3, 0.5, 0.7,0.8 (three flavor). With a fixed negative pressure of vacuum: $B= ( 100 {\rm MeV} )^4$, the stiffness of EOS increases along with $\alpha$.}
	\label{fig:EOS_3fla}
\end{figure}

To discuss the stability, the pressure as a function of quark chemical potential is plotted in Fig. \ref{fig:compare}
and then the EOSs with different parameters of both two-flavor and three-flavor quark matter are separately plotted in Fig. \ref{fig:EOS_2fla} and Fig. \ref{fig:EOS_3fla}. 
In our model, when the bag constant, $\alpha$ and the u quark chemical potential are the same, two-flavor quark matter has a higher pressure than three-flavor quark matter, or, at the same pressure, the two-flavor matter has a lower chemical potential, which is different from what Witten expects in his paper \cite{witten}. Witten estimates the relation between quark chemical potential and pressure based on the MIT bag model and concludes that at the same pressure the quark chemical potential of two-flavor quark matter is higher than that of three-flavor quark matter, which means the particles on the Fermi surface of three-flavor quark matter have lower energy and thus more stable. Besides, comparing the pressure of two possible phases at the same chemical potential is a standard way in thermal physics to determine which phase should exist. Therefore, the two-flavor quark matter is more stable than the three-flavor quark matter in our model.
However, it can also be seen from the Fig. \ref{fig:compare} that the pressure is influenced by $\alpha$ and other parameters, and there is no physical basis that our two-flavor model should has the same $\alpha$ as our three-flavor model. They are actually two different models with different sets of parameters. Because the NJL-type models are phenomenological models where parameters such us the coupling constant, the cutoff and the current quark mass are calibrated by fitting hadronic data and Lattice QCD at zero temperature and chemical potential. A two-flavor or a three-flavor model can have several possible parameter sets. So, our result is also model dependent. 

\begin{figure}
	\includegraphics[scale=0.96]{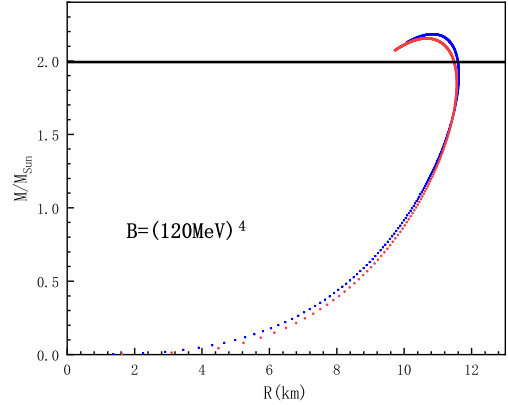}
	\caption{Mass-radii relations for two-flavor and three-flavor condition. The red line and blue line represent the two-flavor condition with $\alpha=0.95$ and three-flavor condition with $\alpha=0.8$ respectively. Masses are scaled by the mass of sun: $M_{sun}$.}
	\label{fig:m_r relation}
\end{figure}

We would like to show that both the two-flavor model and the three-flavor model can satisfy the constraints from astronomical observations. Certainly, the parameters can be constrained by the astronomical observations. At first, in astronomy, the most reliable observation evidence is the existence of two-solar-mass compact stars \cite{demorest2010two,antoniadis2013massive,fonseca2016nanograv,cromartie2019very}. Our mass-radii relations of both two-flavor quark stars and three-flavor quark stars compared with observations are shown in Fig. \ref{fig:m_r relation}.
\begin{figure}
	\includegraphics[scale=0.96]{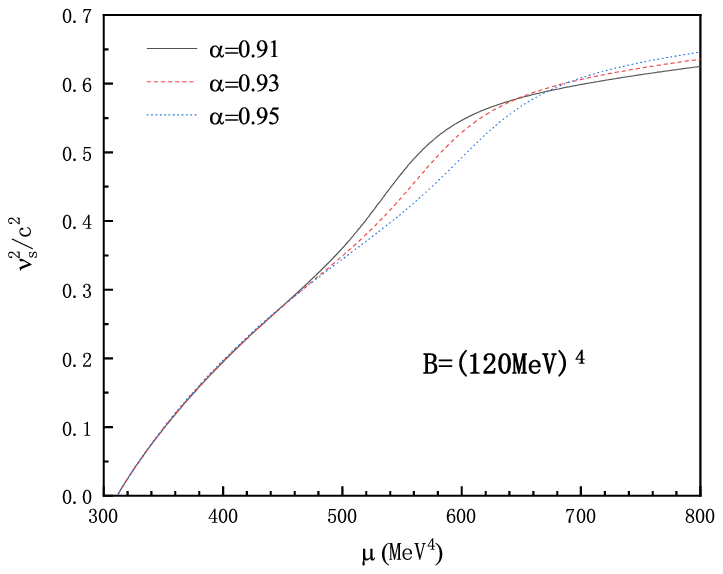}
	\caption{The square of sound velocity as functions of chemical potential of u quark for two-flavor condition. $c$ is the speed of light.}
	\label{fig:vs_2fla}
\end{figure}
\begin{figure}
	\includegraphics[scale=0.96]{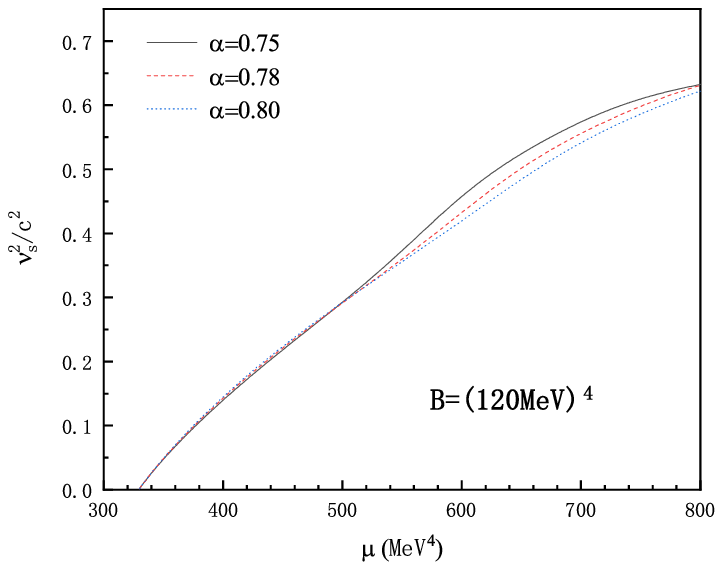}
	\caption{The square of sound velocity as functions of chemical potential of u quark for three-flavor condition. $c$ is speed of light.}
	\label{fig:vs_3fla}
\end{figure}
\begin{figure}
	\includegraphics[scale=0.96]{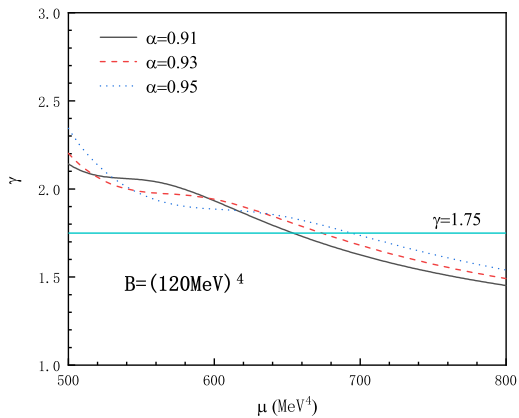}
	\caption{Polytropic index as functions of chemical potential of u quark for two-flavor condition.  The chemical potentials correspond to maximum mass for $\alpha=0.91, 0.93, 0.95$ are 734 MeV, 744 MeV and 754 MeV respectively.}
	\label{fig:poly_2fla}
\end{figure}
\begin{figure}
	\includegraphics[scale=0.96]{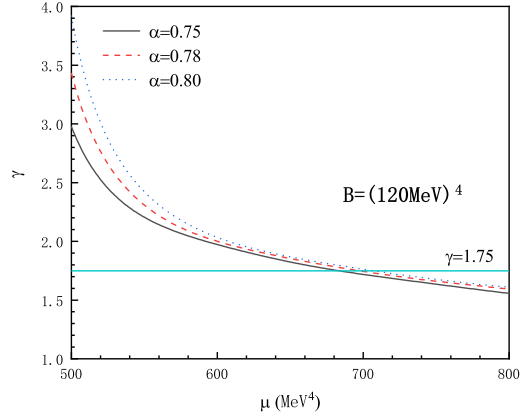}
	\caption{Polytropic index as functions of chemical potential of u quark for three-flavor condition. The chemical potentials correspond to maximum mass for $\alpha=0.75, 0.78, 0.80$ are 765 MeV, 780 MeV and 793 MeV respectively.}
	\label{fig:poly_3fla}
\end{figure}

 If $\alpha$ is large enough, both two-flavor and three-flavor models can construct a two-solar-mass neutron star. From the Lagrangian we can see that the vector-isoscalar channel plays an important role in the Fierz transformed Lagrangian after the mean field approximation is performed, and it is this term that increase the stiffness of EOSs. 
Certainly, it is generally believed that most neutron stars are real neutron stars or hybrid stars and many related works about EOSs of hybrid stars has been done in Refs. \cite{benic2015new,bonanno2012composition}. But the existence of  pure quark stars cannot be ruled out. There are even evidence supporting the existence of pure quark star in astronomical observation \cite{dai2016most}. The two-solar-mass neutron star can also be a quark star. Even if it is not, a massive hybrid star suggests a very stiff equation of state for quark matter \cite{ozel2006soft}. Then, according to our EOSs, for two-flavor quark star with $\alpha=0.95$ and $B$=(120 MeV$)^4$, the tidal deformability for $1.4$ $M_{sun}$ is 557 and the value for three-flavor quark star with $\alpha=0.8$ and $B$=(120 MeV$)^4$ is 570. Both of them can satisfy the constraints determined in the LIGO/Virgo publication that the tidal deformability for $1.4 M_{sun}$ should be below 580 \cite{GW170817PhysRevLett.121.161101}. Finally, the sound velocity of EOSs should be less than the speed of light so that the law of causality will not be broken. The results are shown in Fig. \ref{fig:vs_2fla} and Fig. \ref{fig:vs_3fla}. We also compute the polytropic index ($\gamma=\frac{d(\ln P)}{d (\ln \epsilon})$) as functions of chemical potential which can be employed to analyse the phase structure of QCD matter \cite{annala2020evidence,PhysRevC.102.025203,PhysRevD.102.023021,ma2020composition}. The results are shown in Fig. \ref{fig:poly_2fla} and Fig. \ref{fig:poly_3fla}. The values of $\alpha$ and $B$ we choose for the calculation of speed of sound and polytropic index all satisfy the constraints from mass and tidal deformability. It is interesting to find that our polytropic index for maximally massive stars are all less than 1.75. It is in accord with the criterion suggested in Ref. \cite{annala2020evidence} to judge the existence of quark matter in neutron stars. What we need to point out here is that the above calculation is just an example to show how $\alpha$ can be constrained by astronomical observations. For example, we set $B$=(120 MeV$)^4$ at first. Then we would compute the star with maximum mass and tidal deformability. For two-flavor quark star, when $\alpha$ is smaller than 0.91, we cannot construct a two-solar-mass quark star and if $\alpha$ is larger than 0.95, the tidal deformability will exceed 580. So, under this condition, the $\alpha$ should lie in the interval of 0.91-0.95. Of course, at present, we cannot determine $\alpha$ and $B$ exactly. We need more data of astronomical observations to determine values of them in the future.

\section{\label{sec:discussion}Summary and discussion}
In this paper, we utilize an NJL-type model to study the phase diagram of two-flavor and three-flavor quark matter based on the Pauli-Villars cutoff scheme. A parameter $\alpha$ that can change the position of chiral phase transition point is introduced by our self-consistent mean field approximation. Our phase diagrams have no critical end point. Then, the equation of states of both two-flavor quark matter and three-flavor quark matter are calculated. We find that the two-flavor quark matter is more stable than the three-flavor quark matter in our model, which is different from what Witten expects based on the MIT bag model, but in accordance with a recent work \cite{holdom2018quark}. This result indicates the significance of the interaction between particles in the study of quark stars. So, whether the three-flavor quark matter is more stable than the two-flavor quark matter is model dependent. Up to now there is no reliable way to give the answer. Finally, we obtain the mass-radius relation of quark stars that can be compared with observations to optimize the parameters in our model. These results can also be compared with our previous works with different regularization schemes \cite{zhao2019does,wang2019non}. It is interesting to find that no matter what kind of regularization scheme is used, no matter whether the quark matter is two-flavor or three-flavor, the $\alpha$ should be about 0.85 or much larger to construct a two-solar-mass quark star and the $\alpha$ is almost independent of regularization schemes when constrained by astronomical observations. This conclusion reflect a basic fact: we must increase the intensity of vector-isoscalar channel to satisfy the result of astronomical observations. Otherwise, if we adopt the dense strong interaction equation of state introduced by the normal NJL model, we cannot obtain the two-solar-mass neutron star indicated by the astronomical observation.

\acknowledgments
This work is supported in part by the National Natural Science Foundation of China (under Grants No. 11535005 and No. 11690030), the National Major state Basic Research and Development of China (Grant No. 2016YFE0129300).

\bibliography{reference.bib}
\end{document}